\documentclass[aps,pra,twocolumn,showpacs,groupedaddress]{revtex4}
\usepackage{graphicx}
\usepackage{epsfig}

\def\Ket#1{\left|#1\right>}
{\catcode`\|=\active
  \gdef\Braket#1{\left<\mathcode`\|"8000\let|\bravert {#1}\right>}}
\def\bravert{\egroup\,\vrule\,\bgroup}

\begin{document}

\title{Dynamics of Fermionic Four-Wave Mixing}
\author{H. Christ, C. P. Search, and P. Meystre}
\affiliation{Optical Sciences Center, The University of Arizona,
Tucson, AZ 85721}
\date{\today}

\begin{abstract}
We study the dynamics of a beam of fermions diffracted off a
density grating formed by fermionic atoms in the limit of a large
grating. An exact description of the system in terms of
particle-hole operators is developed. We use a combination of
analytical and numerical methods to quantitatively explore the
Raman-Nath and the Bragg regimes of diffraction. We discuss the
limits in diffraction efficiency resulting from the dephasing of
the grating due to the distribution of energy states occupied by
the fermions. We propose several methods to overcome these limits,
including the novel technique of ``atom echoes''.
\end{abstract}

 \maketitle

\section{Introduction}
The availability of quantum-degenerate bosonic atomic systems has
recently allowed the extension of atom optics to the nonlinear
\cite{Lenz93} and the quantum regimes. Matter-wave four-wave
mixing \cite{Deng99}, coherent matter-wave amplification
\cite{amplification} and superradiance \cite{Inou99a}, the
generation of dark \cite{Burg99} and bright \cite{Stre02} atomic
solitons and of correlated atomic pairs \cite{Voge02} have been
demonstrated, and so has the matter-wave analog of second-harmonic
generation, the creation of a molecular condensate component
\cite{Wyna00,Mcke02}.

In contrast, the development of the nonlinear atom optics of
fermionic atomic systems is not nearly as far along. While it has
been shown theoretically \cite{Kett01,Moor01} that the four-wave
mixing of fermionic matter waves is possible in principle, these
predictions have not been verified experimentally so far. Still,
the recent achievement of temperatures as low as $0.2T_F$, where
$T_F$ is the Fermi temperature, for the fermions $\mbox{ }^{40}$K
and $\mbox{ }^{6}$Li \cite{Lithium1,Dema99,Hadz02,Roat02} is
encouraging, and it is hoped that first experiments on fermionic
nonlinear atom optics will take place in the near future. In
addition to the fundamental goal of achieving a BCS phase
transition into pairing and superfluidity \cite{Thomas02},
research along these lines is also motivated by recent results
that hint at the possibility to lower the phase noise in
interferometric measurements below the bosonic standard quantum
limit by using instead degenerate fermionic beams \cite{Sear02c}.

The first theoretical discussions of fermionic nonlinear atom
optics were presented in Refs. \cite{Kett01} and \cite{Moor01},
which treated the case of a single `test' particle scattering off
a periodic density grating formed by a degenerate Fermi gas. They
showed that for an appropriately prepared grating, the fermionic
system can indeed undergo four-wave mixing. In contrast to the
standard interpretation in terms of ``bosonic amplification'',
which clearly is not applicable to fermions, this effect was
interpreted in terms of the constructive quantum interference
between different paths leading to the same final state.

One important aspect of the fermionic case is that, in contrast to
bosons, considerable care must be taken in combining two matter
waves to form a ``grating'', so that their interaction with a
third wave can produce a significant four-wave mixing signal.
Consider, as we shall do in this paper, counterpropagating matter
waves separated in momentum by $q$. In the case of bosons, two
obvious possibilities correspond to the states
\begin{equation}
|\psi_{b,1}\rangle=\frac{1}{\sqrt{2 (N/2)!}}\left [
(b^\dagger_{q/2})^{N/2}+(b^\dagger_{-q/2})^{N/2} \right ]
|0\rangle,
\end{equation}
and
\begin{equation}
|\psi_{b,2}\rangle=\frac{1}{\sqrt{2^N N!}} \left
[b^\dagger_{q/2}+b^\dagger_{-q/2} \right ]^N|0\rangle ,
\end{equation}
the $b^\dagger_k$ being usual bosonic creation operators and
$|0\rangle$ the atomic vacuum. The fist case describes two
counterpropagating beams of $N/2$ atoms each and of momenta $\pm
q/2$, while the second state corresponds to a density grating
obtained by identically preparing $N$ atoms in a coherent
superposition of states of momenta $\pm q/2$. It is known from the
study of atomic diffraction by optical fields that these two
states lead to different diffraction patterns, because the first
one contains ``which way'' information while the second doesn't
\cite{Shore91}. This difference becomes however insignificant for
large gratings.

The situation for fermions is more complicated, since the Pauli
exclusion principle precludes one from placing more than one atom
per state. One needs instead to consider multimode atomic beams,
centered around the mean momenta $\pm q/2$. In this case the
states $|\psi_{b,1}\rangle$ and $|\psi_{b,2}\rangle$ are replaced
by
\begin{equation}
|\psi_{f,1}\rangle=\frac{1}{\sqrt{2}} \left [\prod_{\{k\}}^{N/2}
a_{k+q/2}^\dagger + \prod_{\{k\}}^{N/2} a_{k-q/2}^\dagger \right
]|0\rangle
\end{equation}
and
\begin{equation}
|\psi_{f,2}\rangle=\prod_{\{k\}}^N  \frac{1}{\sqrt{2}} \left
[a_{k+q/2}^\dagger + a_{k-q/2}^\dagger \right ]|0\rangle ,
\label{fermi-grating}
\end{equation}
where $a_{k\pm q/2}^\dagger$ are fermionic creation operators for
atoms of momenta in the vicinity of $\pm q/2$, the total number of
atoms involved being indicated in the appropriate products. From
Refs. \cite{Kett01,Moor01,Yamamoto}, we know that it is the
quantum coherence apparent in matter-wave states of the form
$|\psi_{f,2}\rangle$ that is responsible for fermionic four-wave
mixing. In order for the required quantum interference to occur,
it is essential that every atom be in a coherent superposition of
momentum states centered around $q/2$ and $-q/2$.

\begin{figure}
\begin{center}
\includegraphics*[width=8cm,height=5cm]{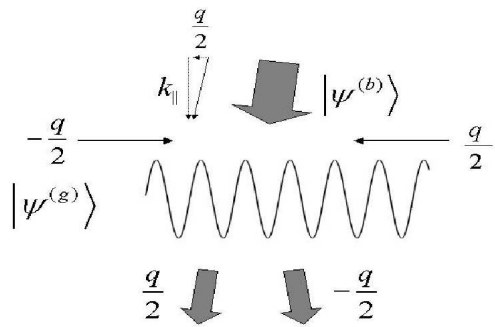}
\vspace{.3 cm}
 \caption{Spin-down polarized fermionic beam diffracted by a second
 species of spin-up polarized fermions forming a matter-wave grating.}
  \label{model}
\end{center}
\end{figure}

So far, our discussion has ignored the time dependence of the
grating. But since the atoms forming a fermionic grating all have
slightly different kinetic energies, their free evolution results
in a dephasing that is expected to eventually lead to the
disappearance of the four-wave mixing signal. Although the
importance of this dephasing was pointed out in Ref.
\cite{Kett01}, no quantitative discussion of its effect has been
presented so far. The present paper addresses this problem
quantitatively by a combined analytical and numerical study of the
diffraction of a beam of fermionic atoms off a large fermionic
grating. We fully include the dynamics of the atomic beam, but
neglect its back-action on the grating dynamics, considering only
its free evolution. This is the matter-wave analog of the
undepleted pump approximation in nonlinear optics.

Section II introduces our model and describes its dynamics in
terms of equations of motion for particle-hole operators. The
effect of the grating dynamics is characterized in terms of a
dephasing time $\tau_d$, whose impact is then illustrated in the
simple case of Raman-Nath diffraction. The Bragg regime is
analyzed in section III using a combination of an analytical
discussion of a simple two-mode system and full numerical
simulations. We determine the characteristic time scales governing
this process, and conclude that four-wave mixing in degenerate
Fermi gases should barely be observable. Noting that the dephasing
of the grating is in principle reversible, we turn in Section IV
to a discussion of possible ways to achieve such a reversal, based
on analogies with the photon echo techniques of quantum optics.
Since the physical origin of the dephasing is the difference in
kinetic energies for atoms with different momenta, one elegant
approach exploits the interaction of the grating with a periodic
external potential to achieve a negative effective mass of the
atoms. Such methods pave the way for the observation of phenomena
that seem out of experimental reach at first glance. Finally,
Section V is a summary and outlook.

\section{Model}
A typical atomic four-wave mixing experiment involves three input
wave packets that interact, typically via $s$-wave collisions, to
produce a fourth packet in a new momentum state. Physically, this
can be interpreted as the result of the phase-matched scattering
of one of the wave packets off the matter-wave grating produced by
the superposition of the other two, as discussed in the
introduction.

As is readily apparent from Eq. (\ref{fermi-grating}), though, a
grating formed by $N$ fermions occupies $2N$ momentum states, with
$N$ states around the momentum $q/2$ and $N$ states near $-q/2$ in
our case. As a result of the free evolution associated with the
spread in kinetic energies of these atoms, the density grating
undergoes a dephasing, similar to Doppler broadening in a thermal
vapor. The goal of this paper is to investigate the effects of
this dephasing and to determine to what extent it can be reduced,
or even eliminated.

We consider the situation of a spin-polarized fermionic beam
diffracted by an atomic grating of spin-polarized fermions that
are either of a different species or in a different spin state,
see Fig.~\ref{model}, the beam and grating atoms interacting
coherently via elastic $s$-wave scattering. We neglect $p$-wave
scattering, whose cross-section at zero temperature is orders of
magnitude smaller than that of \emph{s}-wave scattering.

The diffraction of a fermionic atomic beam by a periodic potential
was previously treated for the case of free fermions (or
equivalently a one-dimensional bosonic Tonks gas) \cite{Rojo99}
and of Cooper pairs \cite{Sear02a}. These treatments are
restricted to the Raman-Nath regime, where the kinetic energies of
the diffracted atoms are negligible compared to the potential
energy of the grating. The scattering of two fermions by an
optical standing wave was also considered in Ref.~\cite{Lenz93b}.
Also worth mentioning in the present context is Ref.
\cite{Vill01}, which investigates the nonlinear mixing of two
bosonic and one fermionic wave. In these cases Pauli blocking has
a significant effect on the diffraction pattern, but as the
gratings are non-fermionic, the dephasing, which is central to our
work, is not addressed.

We restrict the quantized description of the system to the
dimension along the axis of the grating only, the ``transverse''
direction. The dynamics of the system in the direction
perpendicular to it is treated classically, and parameterized by
the time the atoms in the beam interact with the grating. Note
that this approximation neglects both the normal reflection and
the Bragg reflection of the incident beam off the grating
\cite{Vill01}.

This model is described by the second-quantized Hamiltonian
\begin{eqnarray}\label{BasicHamiltonian}
  H&=&\sum_k \left (\hbar \omega_k^{(g)} {a}^{\dag}_{k \uparrow}a_{k
  \uparrow} + \hbar \omega_k^{(b)}{a}^{\dag}_{k \downarrow}a_{k
  \downarrow} \right) \nonumber\\
  &+& \hbar U_0
 \sum_{k_1,k_2,q}\left(
  a^{\dag}_{k_1+q \uparrow}a^{\dag}_{k_2-q \downarrow} a_{k_2
  \downarrow}a_{k_1 \uparrow} \right),
\end{eqnarray}
where $\omega_k^{(g)}=\hbar k^2/2m_g$ ($\omega_k^{(b)}=\hbar
k^2/2m_b$) is the kinetic energy of a grating (beam) atom of mass
$m_g$ ($m_b$) and transverse momentum $k$,
\begin{equation}
U_0= 2\pi \hbar a/(LA\mu), \label{U0}
\end{equation}
$\mu = m_gm_b/(m_g + m_b)$ is the reduced mass, $a$ is the
$s$-wave scattering length characterizing the ultracold collisions
between grating and beam atoms, $A$ is an effective transverse
cross-sectional area, and $L$ is the quantization length. The
annihilation and creation operators $a_{k,s}$ and
$a^\dagger_{k,s}$ satisfy the fermionic anticommutation relations
$\left[ a_{ks}, a_{k's'} \right]_+=0$ and $\left[ a_{ks},
a^{\dag}_{k's'} \right]_+=\delta_{kk'}\delta_{ss'}$, the spin
components $s=\{\uparrow, \downarrow \}$ corresponding to the different spin
states or atomic species of the grating and the beam, respectively.

Since the elementary process underlying matter-wave diffraction
consists in annihilating an atom of momentum $k$ and creating an
atom with momentum $k + q$, it is useful to describe the dynamics
of the atomic beam in terms of ``particle-hole'' operators
\begin{eqnarray}\label{particleholeoperators}
  \rho_{k,q}^{(b)}&=&a^{\dag}_{k+q \downarrow}a_{k \downarrow},
  \nonumber\\
  \rho_{k,q'}^{(g)}&=&a^{\dag}_{k+q \uparrow}a_{k \uparrow},
\end{eqnarray}
where the subscripts ``b'' and ``g'' label beam and grating atoms,
respectively. The Heisenberg equations of motion for these
operators are easily found as
\begin{eqnarray}\label{particleholefullequations}
&&i \frac{d}{dt}\rho_{k,q'}^{(b)}=
\omega_{k,q'}^{(b)}\rho_{k,q'}^{(b)}\nonumber\\
&+&U_0 \sum_{k_1,k_2}\left(
\rho_{k+k_1,q'-k_1}^{(b)}-\rho_{k,q'-k_1}^{(b)}
\right)\rho_{k_2,k_1}^{(g)},
\end{eqnarray}
with $\omega^{b}_{k,q'}=\omega^{b}_k - \omega^{b}_{k+q'}$. A
similar equation holds for the grating particle-hole operators,
with the substitution $b \leftrightarrow g$.

We consider a beam of $N_b$ atoms propagating toward the grating
with some central transverse momentum $\overline{q}$. As discussed
in the introduction the grating is taken to be in a superposition
of states of transverse momenta $k-q/2$ and $k+q/2$, with
$-k_{F,g}\leq k\leq k_{F,g}$, where $k_{F,g}=\pi N_g/L$ is the
Fermi momentum (in one dimension) and $N_g$ is the number of atoms
forming the grating. Such a grating can be prepared with
two-photon Bragg pulses of bandwidth larger than the Fermi
frequency $\hbar k^2_{F,g}/(2m_g)$ in a two-step process starting
from a stationary homogeneous gas, see e.g. Ref. \cite{Tori00}. In
that scheme, a $\pi$-pulse imparting a transverse momentum $-\hbar
q/2$ to each atom in the static cloud is followed by a $\pi/2$
Bragg pulse that creates the desired superposition state. Thus the
initial state of beam-grating system is
\begin{equation}
|\Psi(t=0)\rangle=|\Psi^{(b)}\rangle |\Psi^{(g)} \rangle,
\end{equation}
where
\begin{eqnarray} \label{initialconditions}
 &&\Ket{\Psi^{(b)}}= \prod_{|k| \leq k_{F,b}}
 a^{\dag}_{k+\overline{q} \downarrow}\Ket{0}, \nonumber \\
 &&\Ket{\Psi^{(g)}}=\prod_{|k| \leq
 k_{F,g}}  \frac{1}{\sqrt{2}}\left(a^{\dag}_{k-q/2 \uparrow}+a^{\dag}_{k+q/2 \uparrow}\right)
 \Ket{0}.
\end{eqnarray}

The corresponding initial expectation values of the particle-hole
operators are
\begin{eqnarray}
\label{initialvalues} && \Braket{\Psi^{(b)} |\rho_{k,q'}^{(b)}(0)
| \Psi^{(b)}} = \delta_{q',0}\Theta(k_{F,b}-|k-\overline{q}|),
\nonumber\\
&& \Braket{\Psi^{(g)} |\rho_{k,q'}^{(g)}(0) |\Psi^{(g)}}
=\frac{1}{2}\left( \delta_{q',0} + \delta_{q',q}
\right)\Theta(k_{F,g}-|k+\frac{q}{2}|)\nonumber\\
& & +\frac{1}{2}\left( \delta_{q',0} + \delta_{q',-q}
\right)\Theta(k_{F,g}-|k-\frac{q}{2}|).
\end{eqnarray}

In the following we assume that the matter-wave grating is
sufficiently large that one can safely ignore the back-action of
the atomic beam on its dynamics. Mathematically, this amounts to
neglecting the term proportional to $U_0$ in the grating version
of Eq. (\ref{particleholefullequations}). In this limit, the
grating simply undergoes a free evolution fully governed by its
transverse kinetic energy.

Assuming, consistently with the assumption of a large grating,
that the expectation value of products of beam and grating
particle-hole operators can be factorized as
\begin{equation}
\langle \hat{\rho}_{k,q}^{(g)}\hat{\rho}_{k',q'}^{(b)}\rangle
\approx \langle \hat{\rho}_{k,q}^{(g)}\rangle \langle
\hat{\rho}_{k',q'}^{(b)}\rangle,
\end{equation}
making explicit use of the grating initial condition
(\ref{initialvalues}), and introducing the slowly varying
particle-hole operators
$$\rho_{k,q'}^{(b)}(t)=\hat{\rho}_{k,q'}^{(b)}(t)\exp[-i\omega_{k,q'}^{(b)}t],$$
and similarly for the grating, yields then the beam particle-hole
Heisenberg equations of motion
\begin{widetext}
\begin{eqnarray}\label{lin}
\frac{d}{dt} \hat\rho_{k,p}^{(b)}= -ig(t)\left[
\hat\rho_{k+q,p-q}^{(b)}\exp(i\omega_{k,q}^{(b)}t)-
\hat\rho_{k,p-q}^{(b)}\exp(-i\omega_{k+p,-q}^{(b)}t)
-\hat\rho_{k-q,p+q}^{(b)}\exp(i\omega_{k,-q}^{(b)}t)+
\hat\rho_{k,p+q}^{(b)}\exp(-i\omega_{k+p,q}^{(b)}t) \right],
\end{eqnarray}
\end{widetext}
where
\begin{equation}
\label{timedepcoupling} g(t)=\frac{
U_0}{2}\sum_{k_2}\Theta(k_{F,g}-|k_2+q/2|)\exp[-i\omega_{k_2,q}^{(g)}t],
\end{equation}
reduces in the continuum limit, $ \sum_k \rightarrow L/2 \pi \int
dk$, to
\begin{equation}
g(t)=\frac{U_0N_g}{2}\mbox{sinc}(\hbar qk_{F,g}t/m_g).
\label{goft}
\end{equation}
with $\mbox{sinc}(x)=\sin(x)/x$. For times larger than the {\em
dephasing time}
\begin{equation}
\tau_d \simeq \frac {\pi m_g}{\hbar q k_{F,g} }, \label{taud}
\end{equation}
we have that $g(t) \simeq 0$, that is, the diffraction of the beam
comes to a stop. This consequence of the dephasing between
different momentum states of the atomic grating resulting from
their free evolution, represents an essential difference between
fermionic and bosonic four-wave mixing: the free evolution of a
bosonic grating optically prepared from a Bose-Einstein condensate
at zero temperature gives rise to a phase factor that can easily
be transformed away. For fermions, though, it results in a
dephasing and the effective shutting off of its interaction with
the atomic beam.

This impact of the dephasing is particularly easy to see in the
Raman-Nath regime of atomic diffraction. We recall that this is
the regime where the kinetic energy of the atoms in the beam is
small compared to the potential energy of the density grating.
However, we retain the kinetic energies of the grating atoms,
since they are responsible for the dephasing. As such, this model
corresponds to the case where the atoms in the incident beam are
much heavier than the atoms in the grating.

\begin{figure}
\begin{center}
\includegraphics[width=8cm,height=5cm]{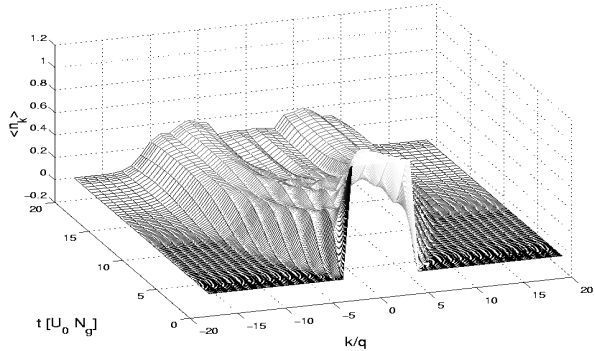}
\vspace{.3 cm}
 \caption{Time evolution of the transverse mode populations of the
diffracted beam, $\langle \hat{n}_k \rangle =\langle
\hat{\rho}_{k,0} \rangle$, in the Raman-Nath regime. The initial
momentum spread of the beam is $k_{F,b}=3 q$. The effects of the
dephasing grating become apparent. Time in units of the inverse
mean-field interaction strength $U_0 N_{g}$ and $\tau_d=15$.}
\label{rn_fermions}
\end{center}
\end{figure}

Approximating the exponential functions by unity in Eq.
(\ref{lin}) consistently with the Raman-Nath approximation gives
$$\frac{d}{dt} \hat\rho_{k,p}^{(b)}=
-i g(t) \left( \hat\rho_{k-q,p+q}^{(g)}+\hat\rho_{k+q,p-q}^{(g)}-
\hat\rho_{k,p-q}^{(g)}- \hat\rho_{k,p+q}^{(g)}\right).
$$
The solution of these operator equations for $g(t)=$ constant are
known \cite{Rojo99} in terms of Bessel functions of the first kind
of order $s$ as
\begin{eqnarray}
\label{ramansol} &&\rho^{(b)}_{k,p}(t)=
\sum_{s,s'=-\infty}^{+\infty}i^{s-s'}J_s(U_0 N_g t) \nonumber\\
& & \times J_{s'}(U_0 N_g t) \rho^{(b)}_{k-sq,sq-s'q+p}(0).
\end{eqnarray}
Due to conservation of momentum, only states of transverse momenta
separated by integer multiples of $q$ are dynamically coupled. For
a beam of initial momentum spread $k_{F,b} < q/2$, this implies
that a linearly increasing sequence of replica of the initial
distribution will be generated in time. Things are more
complicated if $k_{F,b} \ge q/2$, since Pauli blocking leads to a
situation where only those transverse modes near the edge of the
distribution can initially be diffracted, creating holes into
which modes deeper into the initial distribution can then
subsequently be coupled. This results in an effective broadening
of the initial distribution in time, as illustrated in the early
stages of Fig.~\ref{rn_fermions}.

\begin{figure}
\begin{center}
\includegraphics[width=8cm,height=5cm]{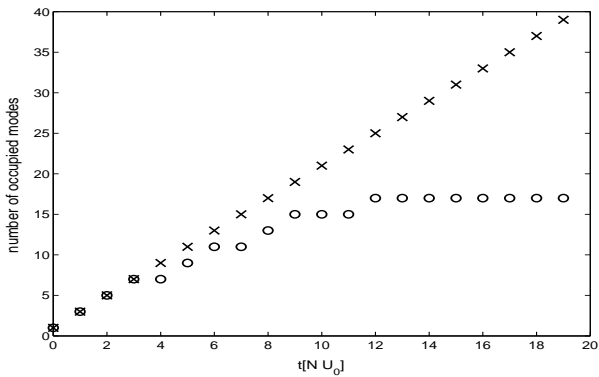}
\vspace{.3 cm}
 \caption{Largest transverse momentum state with a significant population ($>
.03$) as a function of time in the presence (circles) and absence
(crosses) of dephasing. Time is in units of the inverse mean-field
frequency $U_0N_g$ and $\tau_d=15$. } \label{raman_gt}
\end{center}
\end{figure}

Since $g(t)$ is actually not a constant, the solution
(\ref{ramansol}) is only valid for times $t \ll \tau_d$. In
general, it is necessary to solve the expectation value of the
coupled Raman-Nath equations numerically. The effect of the
grating diffusion, and the resulting vanishing of $g(t)$ for long
enough times, are readily apparent in Fig.~\ref{rn_fermions},
which shows that the beam diffraction effectively ceases for $t >
\tau_d$. This is seen even more clearly in Fig.~\ref{raman_gt},
which compares the number of significantly occupied transverse
modes with and without dephasing of the grating.

\section{Bragg Regime}
We now turn to the analysis of the Bragg regime of fermionic
atomic diffraction. We recall that for single atoms and a grating
of period $2\pi/q$, this regime corresponds to the situation where
only two degenerate modes, of transverse momenta $\pm q/2$, are
coupled. The probabilities of occupation of these two states
oscillate periodically in time, the so-called
\emph{Pendell\"osung} \cite{Bern81}. Bragg scattering is widely
used in atomic beam splitters, both for single atoms \cite{Mart88}
and for Bose-Einstein condensates \cite{Kozu99}.

Fermi statistics complicate the situation in two ways: First, as
we have seen, the distribution of momentum states in the grating
around $\pm q/2$ leads to a dephasing resulting in the
time-dependent coupling constant $g(t)$. In addition, only those
fermions with initial transverse momentum sufficiently close to
$q/2$ fulfill the Bragg condition. Hence, it is expected that
Bragg diffraction will burn a spectral hole in the initial Fermi
sea of the beam, leaving atoms away from that hole practically
untouched.

We can gain some insight into the effects of the dephasing by
considering a simple two-state model that retains only the
coupling between the transverse modes $\pm q/2$, neglecting their
coupling to higher-order modes. This approximation is justified if
the phase-matching condition is strongly violated for these modes,
in other words, if the difference in kinetic energy
$\omega^{(b)}_{q/2,q}$ is much larger than the mean-field energy
$\hbar U_0 N_g$ responsible for mode coupling, i.e.,
\begin{equation}
q^2 \gg 4\pi a \rho_g\frac{m_b}{\mu},
\end{equation}
$\rho_g$ being the atomic density in the grating.

This situation is described by the closed set of operator
equations
\begin{eqnarray}
\frac{d}{dt}\hat{n}^{(b)}_{\pm q/2}&=&\pm ig(t)\left[
\hat{\rho}^{(b)}_{-q/2,q}+\mbox{c.c.}
\right],\nonumber\\
\frac{d}{dt}\hat{\rho}^{(b)}_{-q/2,q}&=& ig(t) \left[
\hat{n}^{(b)}_{q/2}- \hat{n}^{(b)}_{-q/2}\right],
\end{eqnarray}
where $g(t)$ is given by Eq. (\ref{goft}). These equations are
reminiscent of the optical Bloch equations that describe the
dynamics of a two-level atom driven by a resonant single-mode
field of time-dependent amplitude, with the operators ${\hat n}$
corresponding to the populations of the two levels and the
particle-hole operator $\hat{\rho}^{(b)}_{-q/2,q}$ to the atomic
polarization.  The evolution of their expectation values can
readily be found for the initial conditions
$\langle\hat{n}^{(b)}_{q/2}\rangle=1$ and
$\langle\hat{n}^{(b)}_{-q/2}\rangle=0$ as
\begin{equation}
\label{analytical} \langle\hat{n}^{(b)}_{\pm q/2}\rangle\left(
t\right)=\frac{1}{2}\left(1 \pm \cos\beta(t) \right),
\end{equation}
where
\begin{equation} \label{beta}
\beta(t)=\int_0^t dt' g(t') = \frac{1}{2}N_{g}U_0 \int_0^t dt'
\mbox{sinc}\left( \frac{\hbar k_{F,g}q}{m_g}t'\right).
\end{equation}
Similarly to the situation that we encountered in the Raman-Nath
regime, the dephasing of the grating results in the diffraction
effectively coming to an end after a time of the order of
 $\tau_d$.
\begin{figure}
\begin{center}
\includegraphics[width=8cm,height=5cm]{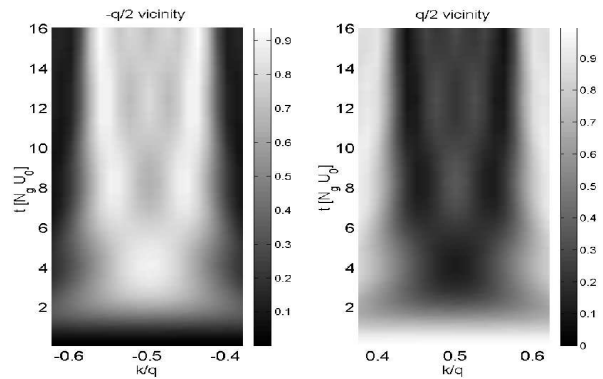}
\vspace{.3 cm}
 \caption{Time evolution of the transverse mode populations of a fermionic
$\mbox{ }^{6}$Li-beam scattering off a $\mbox{ }^{40}$K-grating in
the Bragg regime. Time is in units of the inverse mean-field
frequency $N_g U_0$. In this example $\tau_d\approx 4.2/(N_{g}
U_0)$ and $\omega^{(b)}_q/N_gU_0=5$. The full momentum widths of
the grating and of the beam are $1.0q$ and $0.25q$, respectively.}
  \label{bwscatter}
\end{center}
\end{figure}

Fig.~\ref{bwscatter} summarizes the results of the numerical
integration of the full equations of motion
(\ref{particleholefullequations}) in the Bragg regime for a
$\mbox{ }^{6}$Li beam scattering off a grating formed by $\mbox{
}^{40}$K atoms. It shows the population of the transverse modes of
the atomic beam as a function of time for the case where the
initial full transverse momentum width of the grating and beam are
$1.0 q$ and $0.25 q$, respectively. The hole burned in the initial
Fermi sea is clearly visible, as is the impact of the dephasing of
the grating. The first maximum in the Pendell{\"o}sung
oscillations is strongly pronounced for modes near the phase
matched mode $q/2$, but further maxima cannot be observed due the
dephasing of the fermionic grating.

\begin{figure}
\begin{center}
\includegraphics[width=8cm,height=5cm]{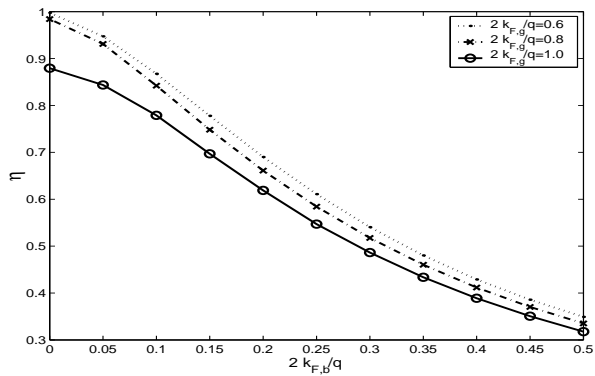}
\vspace{.3 cm}
 \caption{$\eta$ as a function of beam size, $k_{F,b}$, for different sized
 gratings, with $\omega^{(b)}_q/N_gU_0=5$.
 The efficiency $\eta$ decreases with separately increasing
 size of the Fermi seas of beam and grating.
 Nevertheless for reasonably small momentum spreads a decent
 efficiency is obtained. }
  \label{eff_dephase}
\end{center}
\end{figure}

One can characterize the ``efficiency'' $\eta$ of the four-wave
mixing process by the first maximum of the quantity
\begin{equation}
\eta(t)=\sum_{k=-q}^0 \langle \hat{n}_k\rangle\left( t
\right)/N_{b},
\end{equation}
which measures the maximum fraction of atoms whose transverse
momentum has been flipped about $k=0$. It is plotted in
Fig.~\ref{eff_dephase} as a function of the momentum spread of the
incident $\mbox{ }^{6}$Li beam for three widths of the $\mbox{
}^{40}$K grating. The behavior of $\eta$ is as expected, since a
wider incident beam contains more atoms that do not fulfill the
Bragg condition, and a larger Fermi sea of the grating means a
shorter dephasing time.

Experiments are likely to be carried out in a three-dimensional
geometry, in which case the time-dependent coupling constant
(\ref{timedepcoupling}) becomes
\begin{equation}
g_{3D}(t)=3U_0N_{g}\left( \mbox{sinc} \frac{t}{\tau_d}
-\cos\frac{t}{\tau_d} \right)\tau_d^2/t^2,
\end{equation}
where $\tau_d$ is still given by Eq. (\ref{taud}). It is easily
seen that $g_{3D}(t)$ decays in a manner similar to $g(t)$, so
that the lifetime of the grating is essentially the same as in
one-dimension, provided that $k_{F,g}$ remains the same. Note
however that these two times scale differently with the density of
the grating, due to the different mode densities in one and three
dimensions.

In order to assess the experimental feasibility of a fermionic
Bragg diffraction experiment it is necessary to compare the
dephasing time $\tau_d$ to the time $\tau_B$ that it would take to
observe a full Pendell{\"o}sung oscillation in the absence of
dephasing. With the approximation $\beta(t)\approx N_gU_0t/2$ we
have readily that $\tau_B \approx \pi/N_g U_0$. Hence, the
observation of Pendell{\"o}sung oscillations requires
\begin{equation}
\frac{\tau_d}{\tau_B} = \left (\frac{2\pi}{(6\pi^2)^{1/3}}\right
)\left ( \frac{m_g}{\mu} \right )\left (\frac{a
\rho_g^{2/3}}{q}\right )\gg 1, \label{exp-cond}
\end{equation}
where we have used the three-dimensional relationship between the
density $\rho_g$ and $k_{F,g}$, $k_{F,g}=(6\pi^2 \rho_g)^{1/3}$.

Consider for example the diffraction of fermionic potassium
$^{40}$K off a spin-polarized $^{40}$K grating of peak number
density $\rho_g= 10^{15} \mbox{ cm}^{-3}$ and a scattering length
of $85a_0$. Assuming a grating period of 500 nm $\tau_d\approx
0.1\tau_B$, so that Pendell{\"o}sung oscillations will not be
observable in that case. The next section discusses possible ways
around this difficulty.

\section{Atom Echoes}
The condition (\ref{exp-cond}) shows that one possible way to
increase the ratio of $\tau_d$ to $\tau_B$ is to increase the
grating density $\rho_g$. This, however, is likely to be
impractical. Another solution is to increase the scattering length
$a$ via a Feshbach resonance. Noting that $m_g/\mu=(m_g+m_b)/m_b$,
we can also gain an order of magnitude or so in the ratio
(\ref{exp-cond}) by diffracting light fermions such as $^6$Li off
a potassium grating. Ketterle \emph{et. al}~\cite{Kett01} also
proposed a direct way to increase the decay time in three
dimensions by reducing the effective Fermi wave vector in the
direction of the grating by increasing the size of the Fermi sea
in the other directions.

An alternative method that allows one to completely reverse the
effects of the dephasing is based on the observation that this
dephasing is coherent, simply resulting from the different kinetic
energies of the atoms forming the grating. This is reminiscent of
Doppler broadening in laser spectroscopy, which leads to a
dephasing of the atomic polarization. In that case, the dephasing
can easily be reversed by photon echo techniques \cite{abel66}.
Things are slightly more complicated here, because the phase of
the individual atoms is just given by their kinetic energy
$\hbar^2 k^2/2m_g.$ The only way to reverse the sign of this phase
is to reverse the mass of the atoms. This can be achieved by
placing the atoms forming the grating in an additional periodic
potential that gives them a {\em negative effective mass}.

\begin{figure}
\begin{center}
\includegraphics[width=8cm,height=5cm]{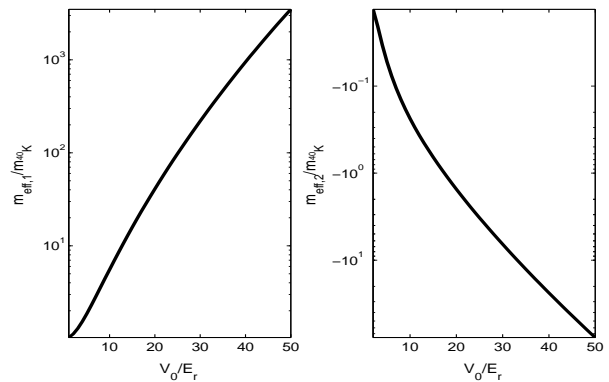}
\vspace{.3 cm}
\caption{ Effective masses of $^{40}$K of the first
two bands at the center of the first Brillouin zone as a function
of $V_0$, in units of the recoil energy $E_r=\hbar^2 k^2 /2 m_g$
for an optical wavelength of 852 nm. }
  \label{effmass}
\end{center}
\end{figure}

Ignoring two-body collisions as well as whatever external fields
are required to prepare the initial grating
(\ref{initialconditions}) for now, the evolution of the grating
atoms in the presence of an optical lattice is governed by the
Hamiltonian
\begin{equation}
H=\frac{\hat{p}^2}{2m}+\frac{V_0}{2}\cos(2k_Lx),
\end{equation}
where $V_0$ is the lattice depth and $k_L$ the wave vector of the
laser creating the periodic potential. The band structure
associated with this potential is readily obtained numerically,
yielding at the center of the first Brillouin zone the effective
mass
\begin{equation}
m_{\rm eff,n} = \hbar^2 \left( \frac{d^2E_n}{dk^2} \right)^{-1}
\Big|_{k=0},
\end{equation}
where $n$ is the band index and $k$ is the quasi-momentum. Figure
\ref{effmass} plots this effective mass of $^{40}$K for the first
two energy bands. As is well known, $m_{\rm eff}$ is alternatively
positive and negative as the band index is increased. We remark
that since $U_0$ is a pseudo-potential adjusted so that the
scattering problem associated with two-body collisions
asymptotically produces the correct collisional cross-section, the
appearance of an effective mass does not change the reduced mass
$\mu$ in Eq. (\ref{U0}). Hence Eq. (\ref{exp-cond}) becomes
\begin{equation}
\frac{\tau_d}{\tau_B} = \left (\frac{2\pi}{(6\pi^2)^{1/3}}\right
)\left ( \frac{m_{\rm eff}}{\mu} \right )\left (\frac{a
\rho_g^{2/3}}{q}\right ).
\end{equation}
From Fig. \ref{effmass}, we see that the effective mass can be
orders of magnitude larger than $m_g$, leading to much easier
conditions to satisfy to observe fermionic Bragg diffraction.

\bigskip

\begin{figure}
\begin{center}
\includegraphics[width=8cm,height=11cm]{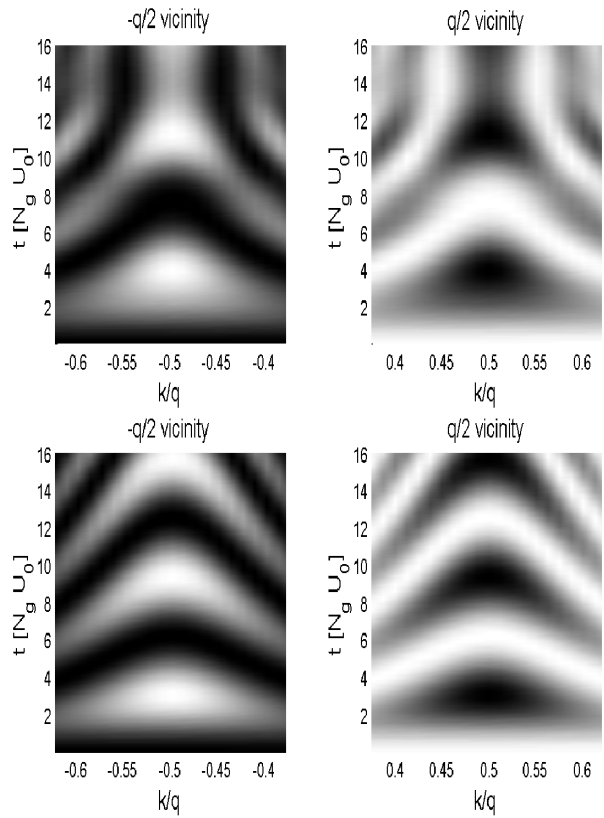}
\vspace{.3 cm}
 \caption{Comparison between the rephasing of the Pendell{\"o}sung
 oscillations by the atom echoes technique (upper plots) to the case where the
 dephasing has been arbitrarily turned off (lower plots). In the plots white
 corresponds to unit probability and black indicates zero probability.
 All parameters as in Fig.~\ref{bwscatter}.}
  \label{compare}
\end{center}
\end{figure}
Even more interesting is the use of the negative effective mass
associated with the second energy band, since it permits a full
reversal of the dephasing. The excitation of the grating atoms to
that band can be achieved e.g. via a two-photon Raman with
co-propagating laser beams, or by modulating the phase of one of
the lattice beams. Assuming for simplicity that the effective
masses corresponding to the $n=1$ and $n=2$ bands are equal in
magnitude, it is easily seen that if the Raman pulse is applied at
time $t_1$, then $g(t_2=2t_1)= g(0)$, and the dephasing has been
completely reversed. Clearly, a sequence of pulses alternatively
transferring the grating atoms between the two energy bands will
keep alternating the sign of $m_{\rm eff}$, thereby continuously
compensating the Doppler broadening. In case the magnitude of
curvatures of the two bands are different, the rephasing time
needs to be appropriately adjusted.

Fig.~\ref{compare} shows the result of this atom echo technique.
The second peak of the \emph{Pendell\"osung} is now clearly
visible, in contrast to the situation in Fig.~\ref{bwscatter}. For
better comparability to the case of uncontrolled dephasing we
considered the same magnitudes for effective and real mass. We
have transferred the atoms to the first band at scaled time $t=3$
and back to the lowest band at $t=9$. The results of the dephasing
of the grating start again later on, but could be eliminated by
transferring the atoms back to the first energy band.

\section{Conclusion}

In this paper, we have investigated the behavior of purely
fermionic scattering, i.e. a beam of fermions diffracted by a
fermionic density grating, in the limit of a large grating. We
explored different regimes which are determined by the relative
magnitude of mean field energy and kinetic energy. Fermi statistic
shows its distinct signatures in the limited lifetime of the
grating and the inability to satisfy the Bragg condition for large
input beams. We proposed novel ways to increase the dephasing time
of the fermionic grating which is the most severe limitation for
fermionic four-wave mixing. Future work will focus on the complex
dynamics arising from the back-action of the scattered atoms on
the grating.

This work is supported in part by the US Office of Naval Research,
by the National Science Foundation, by the US Army Research
Office, by the National Aeronautics and Space Administration, and
by the Joint Services Optics Program. H. C. acknowledges the
support of the Studienstiftung des deutschen Volkes.


\end{document}